\documentclass[fleqn,usenatbib]{mnras}

\usepackage{newtxtext,newtxmath}

\usepackage[T1]{fontenc}
\usepackage{ae,aecompl}

\usepackage{graphicx}	
\usepackage{amsmath}	
\usepackage{amssymb}	
\usepackage[utf8]{inputenc}
\usepackage[normalem]{ulem}

\usepackage{aas_macros}
\usepackage{booktabs}
\usepackage{fullpage}
\usepackage{graphicx}
\usepackage{multirow}
\usepackage{natbib}
\usepackage{subcaption}
\usepackage{siunitx}
   
\newcommand{\code}[1]{\texttt{#1}}
\newcommand{\dirac}{\ensuremath{\delta^\text{D}}}
\newcommand{\lcdm}{$\Lambda$CDM}

\newcommand{\Omm}{\ensuremath{\Omega_\text{m}}}

\newcommand{\diff} {\mathrm{\ensuremath{d}}} 
 
\newcommand{\deriv}[2]{\ensuremath{\frac{\diff {#1}}{\diff {#2}}}}

\newcommand{\mean}[1]{\ensuremath{\left\langle #1 \right\rangle}}
\newcommand{\abs}[1]{\ensuremath{\left\lvert#1\right\rvert}}

\newcommand{\vect}[1]{\ensuremath{\mathbfit{#1}}}
\newcommand{\tens}[1]{\ensuremath{\mathbfss{#1}}}

\newcommand{\hMpc}[1]{\ensuremath{{#1}\,h^{-1}\,\mathrm{Mpc}}}
\newcommand{\Mpch}[1]{\ensuremath{\num{#1}\,h\,\mathrm{Mpc}^{-1}}}
\newcommand{\hMsun}[1]{\ensuremath{\num{#1}\,h^{-1}\,M_\odot}}

\title[Probing the primordial spectrum with the LSS]{Probing features  in the primordial perturbation  spectrum with
  large-scale structure data} 

\author[B.  L'Huillier et al. ]{Benjamin
  L'Huillier,$^{1}$\thanks{E-mail:
    benjamin@kasi.re.kr (BL),
    shafieloo@kasi.re.kr (AS), 
    dhiraj.kumar.hazra@apc.univ-paris7.fr (DKH),
    gfsmoot@lbl.gov (GFS),
    alstar@landau.ac.ru (AAS)}
Arman Shafieloo,$^{1,2}$
Dhiraj~Kumar~Hazra,$^{3,4}$
\newauthor
George~F.~Smoot,$^{5,6}$
and Alexei~A.~Starobinsky$^{7,8}$
\\
$^{1}$Korea   Astronomy   and   Space   Science   Institute,
Yuseong-gu, 776 Daedeok daero, Daejeon 34055, Korea\\
$^{2}$University of  Science and Technology,  Yuseong-gu 217
Gajeong-ro, Daejeon 34113, Korea\\
$^3$AstroParticule et  Cosmologie (APC)/Paris Centre  for Cosmological
Physics, Université Paris Diderot, CNRS, CEA,\\ 
Observatoire de Paris,  Sorbonne Paris Cité University,  10, rue Alice
Domon et Leonie Duquet, 75205 Paris Cedex 13,\\ 
France\\
$^4$Istituto Nazionale di Fisica Nucleare, Sezione di Bologna, Viale Berti Pichat 6/2, I-40127 Bologna, Italy
\\
$^5$Helmut  and Anna  Pao  Sohmen Professor-at-Large,  IAS, Hong  Kong
University of Science and Technology, Clear \\
Water Bay, Kowloon, 999077 Hong Kong, China\\
$^6$Physics  Department  and  Lawrence Berkeley  National  Laboratory,
University of California, Berkeley, 94720 CA, USA\\ 
$^7$L. D. Landau Institute for Theoretical Physics RAS, Moscow 119334,
Russia \\ 
$^8$National Research  University Higher  School of  Economics, Moscow
101000, Russia 
}

\date{Accepted 2018 March 15. Received 2018 February 27; in original form 2017 November 13}

\pubyear{2017}

\begin{document}
\label{firstpage}
\pagerange{\pageref{firstpage}--\pageref{lastpage}}
\maketitle

\begin{abstract}
The form of the primordial power spectrum (PPS) of cosmological scalar
(matter density) perturbations is not yet constrained satisfactorily 
in  spite of  the tremendous  amount  of information  from the  Cosmic
Microwave Background (CMB) data.
While a smooth  power-law-like form of the PPS is  consistent with the
CMB data, some PPS with small  non-smooth features at large scales can
also  fit  the CMB  temperature  and  polarization data  with  similar
statistical evidence. 
Future CMB surveys cannot help distinguish  all such models due to the
cosmic variance at large angular scales.
In this  paper, we study  how well  we can differentiate  between such
featured  forms of  the PPS  not otherwise  distinguishable using  CMB
data.
We ran  15 $N$-body DESI-like  simulations of these models  to explore
this approach.
Showing  that  statistics such  as  the  halo  mass function  and  the
two-point  correlation  function are  not  able  to distinguish  these
models  in a  DESI-like  survey,  we advocate  to  avoid reducing  the
dimensionality  of the  problem by  demonstrating  that the  use of  a
simple three-dimensional count-in-cell density  field can be much more
effective for the purpose of model distinction. 
\end{abstract}




\section{Introduction} 
So  far, the  CMB data  have been  able to  provide us  with the  most
valuable information about the early Universe (EU).  
CMB  temperature anisotropy  and  E-mode polarization  data contain  a
convolved  form  of the  primordial  power  spectrum (PPS)  of  scalar
(matter  density)   perturbations  and  of  the   evolution  of  these
primordial fluctuations through different epochs of the Universe.  

To better understand the physics of  the EU, it is extremely important
to put constraints on the shape of the PPS with high precision.
In particular, looking for small features in the PPS is very important
to  evaluate  whether  it  is  necessary to  go  beyond  the  simplest
theoretical models of the EU.  
For instance, in inflationary cosmology the Universe underwent a phase
of  rapid expansion  known as  inflation in  the remote  past, seeding
structure formation.  
In  its simplest  form, inflation  yields a  power-law-like, close  to
scale-invariant PPS 
\begin{equation}
P_i(k) = A_s \left(\frac k {k_0}\right)^{n_s-1},
\end{equation} 
where the spectral index $n_s$ is close to unity and weakly depends on
the inverse scale  $k$, so it can be approximated  by a constant value
$n_s\simeq 0.96$ over the measured range of scales ($\sim 1-10^4$ Mpc)
\citep{1980PhLB...91...99S,  1981PhRvD..23..347G, 1981JETPL..33..532M,
  1982PhLB..108..389L}.
Thus, the observed PPS is  very well described by only two phenomenological
dimensionless constants.  

However, it is clear that this is the first approximation only, and at
a higher  level of accuracy $\lesssim  10\%$, one can well  expect the
existence of small  non-smooth features in the  PPS whose quantitative
description would  require more phenomenological parameters  (at least
two of them -- their location and relative magnitude).
From the  theoretical point of view,  some new physics will  be needed
for their explanation and derivation.
It  should  be  noted  that  this is  not  specific  for  inflationary
cosmology  only, but  equally well  refers to  any viable  alternative
cosmology of the EU.
It  is simply  that an  alternative cosmology  will require  different
additional physics to predict the same observed feature in the PPS, if
it can accomplish it at all.

In this work we mainly focus on few different inflationary models that
can generate  extra features  to the  power-law shape  of the  PPS and
compare  them with  the  case  of a  power-law-like  form  of the  PPS
predicted by the  simplest inflationary models with  one, maximum two,
free dimensionless parameters fixed by observations.
However, as we  mentioned earlier, the whole analysis  is mainly about
detecting small  non-smooth features in  the PPS,  and thus it  can be
relevant for any alternative EU scenario.

The  study of  the cosmological  microwave background  (CMB), has  put
strong  constrains on  power-law-like  form of  the  PPS predicted  by
simple      single-field       slow-roll      inflationary      models
\citep[e.g.][]{2016A&A...594A..20P}.  
However, degeneracies in the CMB  data allow also several other models
to survive where these models  can generate some prominent features in
the              form              of             the              PPS
\citep[e.g.][]{1992JETPL..55..489S, 2001PhRvD..64l3514A, 2008PhRvD..77b3514J, 2009JCAP...06..028J, 2014JCAP...08..048H, 2014PhRvL.113g1301H, 2017PTEP.2017i3E01H,2018JCAP...02..017H}.  
It is thus important to falsify these  models and test how well we can
distinguish between them using different cosmological observations.  

\begin{figure*}
  \centering
  \includegraphics[width=\textwidth]{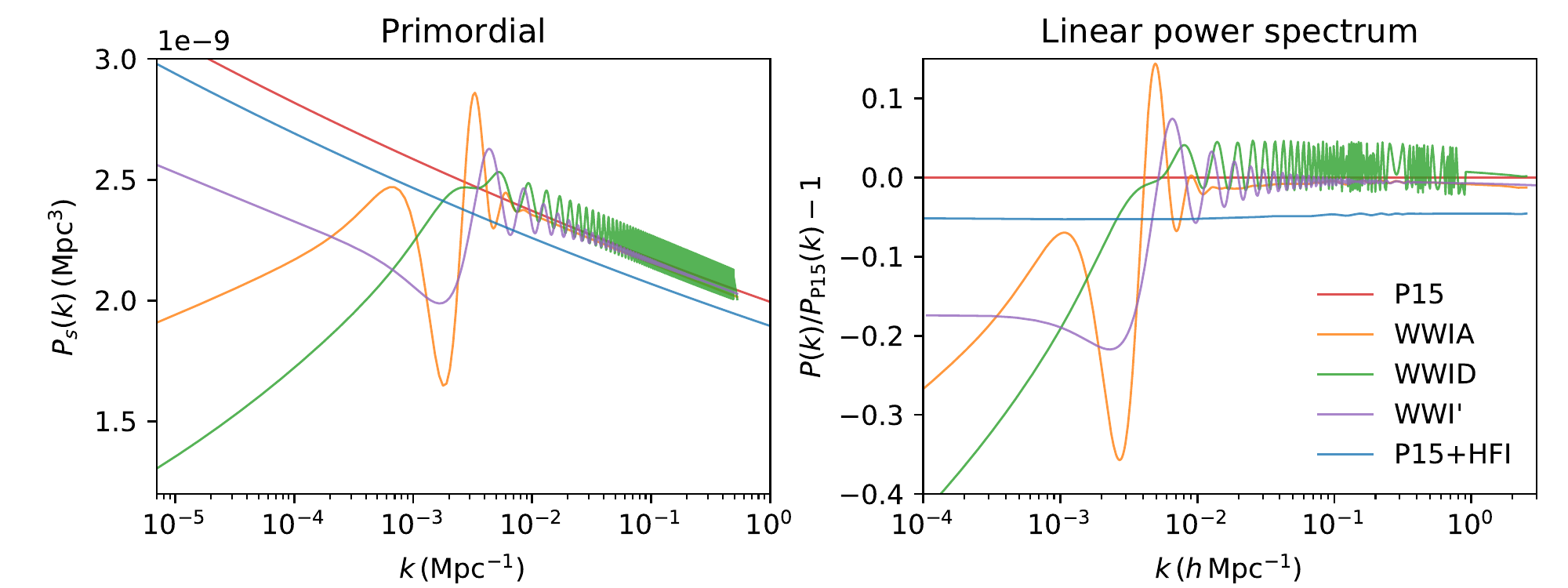}
  \caption{\label{fig:Pkl}
    Left: PPS of our five models.
    Bottom: linear  matter power  spectra normalized by  the reference
    (P15) case. 
  }
\end{figure*}

In  this  work, we  aim  to  assess  the  ability of  the  large-scale
structure (LSS) to differentiate between  few forms of the PPS, result
from some  inflationary models,  that are indistinguishable  from each
other using the CMB data alone.  
Indeed, observationally, one  has two main anchors  that the cosmology
must fit. At  high redshift, the CMB has revealed  a tremendous amount
of information,  strengthening the concordance model  of cosmology and
rising cosmology to a precision science.  
However, even though polarization still holds valuable information, we 
are  about  to  reach  the  limits  of what  can  be  learned  by  the
temperature anisotropy on large scales due to cosmic variance.  
At  low redshift,  the large-scale  structure (LSS)  of the  Universe,
thanks to their three-dimensional structure, also carries a large amount
of information.  Any viable  model must  therefore confront  these two
anchors and fit all the data appropriately.

Upcoming  surveys such  as  the dark  energy spectroscopic  instrument
\citep{2016arXiv161100036D},  EUCLID  \citep{2011arXiv1110.3193L},  or
LSST \citep{2008arXiv0805.2366I} will probe the Universe to an  unprecedented scale and it is important
to know how  well we can use  them to learn more about  the physics of
the early universe.
Cosmological simulations have now  been extensively used to understand
non-linear structure formation \citep[e.g.][]{2005Natur.435..629S} and
test   dark  energy   \citep{2012arXiv1206.2838A,2012MNRAS.422.1028B},
modified   gravity  \citep{2011PhRvD..83d4007Z,   2012JCAP...10..002B,
  2013JCAP...04..029B,    2014PhRvD..89h4023L,   2017MNRAS.468.3174L},
neutrinos     \citep{2014MNRAS.440...75B}, and    warm    dark     matter
\citep{2016MNRAS.457.3492H}. 
In this  work, we  use cosmological  $N$-body simulations,  similar to
what we expect  to observe from DESI, with some  specific forms of the
PPS and  study  how  well we  can  distinguish
between these models.  

Section~\ref{sec:simmod} presents the models and the simulations, the
results are shown in \S~\ref{sec:res}, and we draw our conclusions in
\S~\ref{sec:ccl}.

\begin{table}
  \centering
  \caption{\label{tab:cosmo}Cosmological parameters of models. 
    The last column shows the improvement in $\chi^2$ of the WWI models 
    with respect to  the reference P15 case. Note  that, since P15+HFI
    uses a different 
    data set, we do not compare its $\chi^2$. 
  }
  
  \begin{tabular}{lccccc}
    \toprule
    Model & \Omm & $H_0$ & $\sigma_8$ & $n_\mathrm{s}$& $\Delta\chi^2$\\
    & & (\si{km.s^{-1}} & & &\\
    & &  \si{.Mpc^{-1}}) & & & \\      
    \midrule
    P15 & 0.317 & 67.05 & 0.836 &  0.9625 & 0\\
    WWIA & 0.320 & 66.86 & 0.834 & -& 7\\
    WWID & 0.318 & 67.01 & 0.842 & -& 13.3\\
    WWI$'$ & 0.317 & 67.04 & 0.834 & -& 12\\
    P15+HFI & 0.319 & 66.93 & 0.816 & 0.9619 & -\\
    \bottomrule
  \end{tabular}
\end{table}

\section{Models and simulations}

\label{sec:simmod}
\subsection{The models}

In  this  section, we  present  the  different early  Universe  models
studied in this work. The choice of these models is solely due to the form of the primordial spectrum they generate and the fact that they all can fit the \emph{Planck} CMB data very well. 
Due to the excellent accuracy with which the standard $\Lambda$CDM
6-parameter model fits the \emph{Planck} data with $l>40$, we restrict ourselves
to models which, first, produce {\em localised} features in the PPS and,
second, fit the smooth small-scale part of the PPS outside the features as
well as possible in terms of the PPS slope $n_s\approx 0.965$ measured by
\emph{Planck}. 
Characteristic examples of such models are the Wiggly-wipped
inflationary (WWI) models introduced in \citet[][]{2014JCAP...08..048H,2016JCAP...09..009H}.
These models produce a better fit to the CMB data than the power-law
model,  and are  thus not  currently distinguishable  from each  other
using the CMB data alone.
Our reference model is  the \citet{2016A&A...594A..13P} best-fitting power-law (TTTEEEE+lowTEB+BKP) cosmology (P15).  
We used three different  Wiggly-whipped  inflation models
\citep[WWI,][]{2014JCAP...08..048H,2016JCAP...09..009H},  namely WWIA,
WWID, and WWI$'$.  
Note that the constraints on the WWI models used here are based on the
Planck 2015 
results \citep{2016JCAP...09..009H}. 
Finally, in order to study the effects of a change of cosmology within
the power-law paradigm, we also include the Planck 2015 TTTEEE +
lowTEB + BKP + HFI	(P15+HFI) best-fitting cosmology.
 
WWIA and  WWID provide an  improvement of  $7$ and $13.3$  in $\chi^2$
compared to P15, respectively. 
They are the local best fits obtained from the same
potential having a discontinuity.  
This potential  has four  extra parameters compared  to the  standard slow
roll case.  
WWI$'$, on the  other hand, provides a  $\Delta\chi^2=12$ improvement in
fit with only two extra parameters.  
WWI$'$ has a discontinuity in the derivative of a continuous potential. 
For     a     complete     table     of     likelihood     comparison,
see~\citet{2016JCAP...09..009H}. 
Table~\ref{tab:cosmo} summarized the  best-fitting cosmological parameters
of each model,  and the last column shows the  improvement in $\chi^2$
of the WWI models with respect to P15.  
We aim to study their effects on the large-scale structure and assess
the power of the LSS to constrain these models, complementary to the
CMB.  

We note here again that the main goal of this study is not to focus on the particular models considered here, but rather to try to separate between different featured forms of the PPS that are not distinguishable from the CMB data alone.

\begin{figure*}
  \centering
  \includegraphics[width=\textwidth]{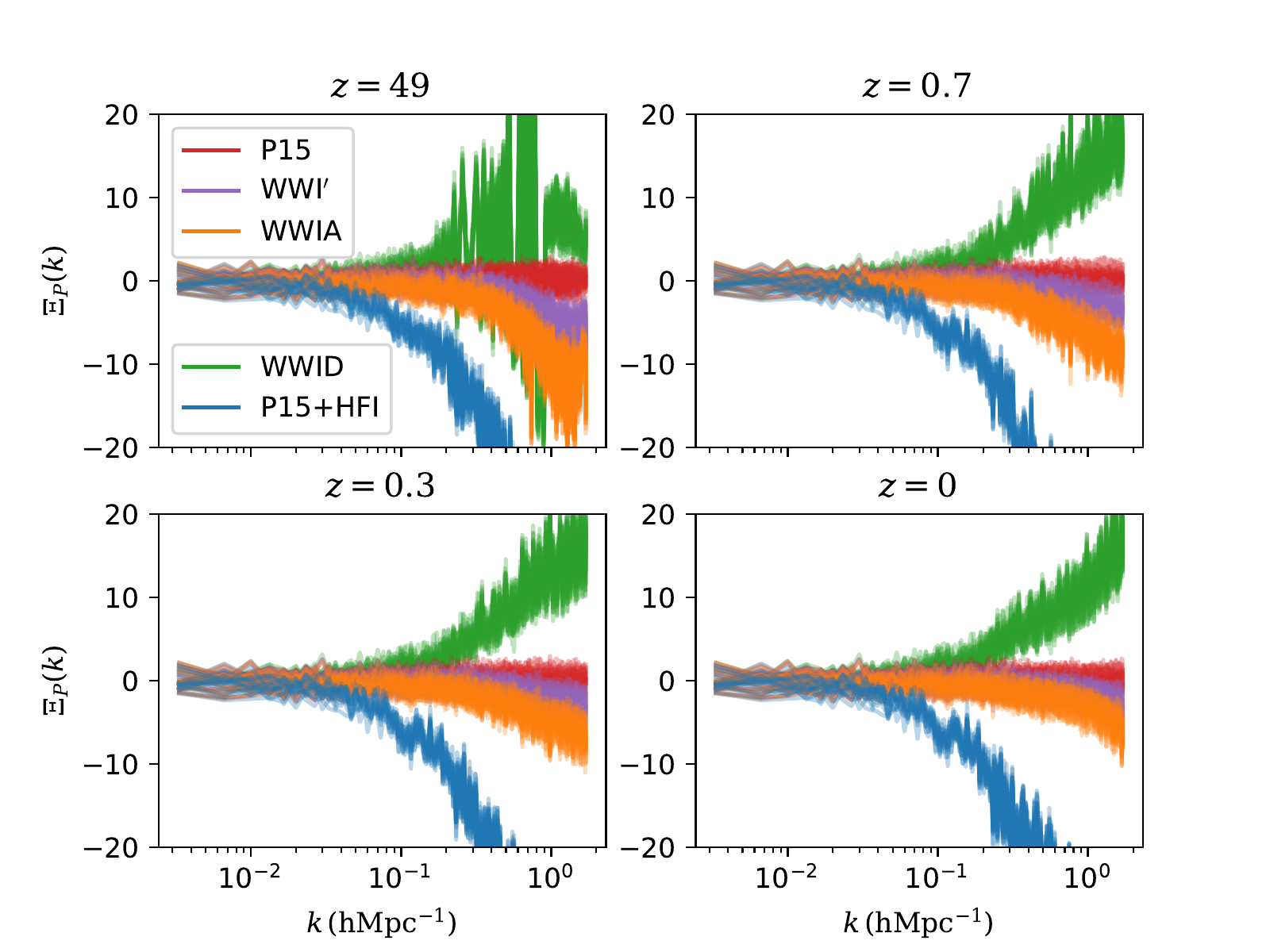}
  \caption{\label{fig:Pk}
    Normalized difference  of the  power spectrum of  each simulation
      with the  average over  the P15  case in  units of  the standard
      deviation of the P15 case
     in
    the initial conditions (top left), and at $z=0.7$ (top right), 0.3 
    (bottom left), and 0 (bottom right). 
  }
\end{figure*}

The  effects of  the  EU  model translate  into  features  in the  PPS
(Fig.~\ref{fig:Pkl}), where the power spectrum of the overdensity field
\begin{align}
  \delta(\vect x) & = \frac{\rho(\vect x)-\bar\rho}{\bar\rho} \\
  \intertext{is given by }
  \mean{\abs{\delta(\vect k) \delta^* (\vect k)}} _ {\abs{\vect{k}}  = 
    k} & = (2\pi)^3 \dirac{(\vect k+\vect{k'})} P(k),
\end{align}
where $\delta(\vect k)$ is the Fourier transform of $\delta(\vect x)$,
and \mean{\cdot} denotes an ensemble average.
In the linear regime, the matter power spectrum can be
calculated from the primordial one by a Boltzmann solver such as
\code{CAMB}\footnote{\url{http://www.cosmologist.info}}
\citep{2000ApJ...538..473L}. 

The  left-hand  panel  of  Fig.~\ref{fig:Pkl} shows  the  PPS  of  the
different models.  
The WWI models essentially results in a superimposition of oscillatory
features on to a smooth power-law. 
We note that these models give indistinguishable CMB angular power
spectra. 
The right-hand panel  shows the linear matter power  spectra  divided
by the reference P15 case.

Most of the difference between the models arises on very large scales
($k\leq\Mpch{e-2}$). 
On these very large-scales, due to the considered volume, the scatter
is very large (few Fourier modes per bin at low-$k$), and the models
cannot be distinguished with upcoming surveys. 
However, the WWI models also show non-power law features at
intermediate scales ($10^{-2}<k<\Mpch{e-1}$). 
Therefore, it is necessary to go to the mildly to fully non-linear
regime, hence the need for cosmological simulations.

\subsection{The simulations}

We    used    the    \code{Gadget-2}    TreePM    cosmological    code
\citep{2005MNRAS.364.1105S}.
The simulation comprises $1024^3$  particles in a \hMpc{1890} box,
assuming a flat-\lcdm\ cosmology. 
The choice of volume is motivated by the expected DESI volume at $z=0.9$ 
with a redshift depth of 0.1 \citep{2016arXiv161100036D,2016arXiv161100037D}. 
The   cosmological   parameters of each model  are   summarized  in
Table~\ref{tab:cosmo}.

The  initial conditions were  generated with  the second-order
Lagrangian perturbation  theory (2LPT)  \code{2LPTic} at  redshift 49,
from a Gaussian random field, since these models have very small
non-gaussianities \citep{2014JCAP...08..048H}. 
2LPT  and high  initial  redshift (for  the  considered mean  particle
separation) yield an accurate power  spectrum and mass function at low
redshift        \citep{1998MNRAS.299.1097S,       2006MNRAS.373..369C,
  2014NewA...30...79L}.

The initial matter  power spectra were obtained at $z=0$   by
\code{CAMB}, and ($z_i=49$) using their own cosmology.
For each model, we chose the best-fitting cosmology given by a modified
version of \code{CAMB}
\citep{2014JCAP...08..048H,2014PhRvL.113g1301H}. 

We note that our $k$-space resolution of $k_0 = 2\pi/L= \Mpch{3.3e-3}$
resolves the  oscillatory features of  the WWI models (the  modes that
are  actually   sampled  are  $k_0  \sqrt{i^2+j^2+l^2},   (i,j,l)  \in
\{0,\dots,N/2\}$).

\section{Results}
\label{sec:res}

\subsection{Power spectrum}

\label{sec:Pk}

In order to estimate the matter power spectrum (PS) in the non-linear
regime, we used the \code{ComputePk} 
code\footnote{Available at 
  \url{http://aramis.obspm.fr/~lhuillier/Codes/ComputePk.php}
}
\citep{2014ascl.soft03015L} to calculate the matter density field in a
grid  with $N_\text{c}^3$  cells, with  $N_\text{c} =1024$,  using the
triangular    shape    cloud     (TSC)    mass    assignment    scheme
\citep{1988csup.book.....H}.

Fig.~\ref{fig:Pk} shows  at a given $k$ 
   	\begin{align}
   	\Xi_{P_i}(k) & = \frac{P_i(k)-\mean{P_\text{P15}(k)}}{\sigma_{P_\mathrm{P15}}(k)},
   	\end{align}
the deviation of the  PS of simulation  of
  each simulation $i$ with respect to  the average of  the PS of  the P15
  case  $\mean{P_\text{P15}(k)}$   in  units  of  the   standard  deviation
  $\sigma_{P_\mathrm{P15}}(k)$ in  the initial conditions (top-left),  and at
  redshifts 0.7 (top-right), 0.3 (bottom-left), and 0 (bottom-right). 
Each line shows one realization of  the PPS, therefore, each model has
15 overlapping lines.

  At  the initial  redshift (upper  left), all  models give  different
  power spectra, and are visually distinguishable. 
On very large scales, due to the  finite volume, only a few modes are
sampled.
Therefore,   $\sigma_{P_\mathrm{P15}}(k)$   is   large  and   the   different
simulations cannot be distinguished from each other.

\begin{figure*}
  \centering
  \begin{subfigure}{.495\textwidth}
    \includegraphics[width=\columnwidth]{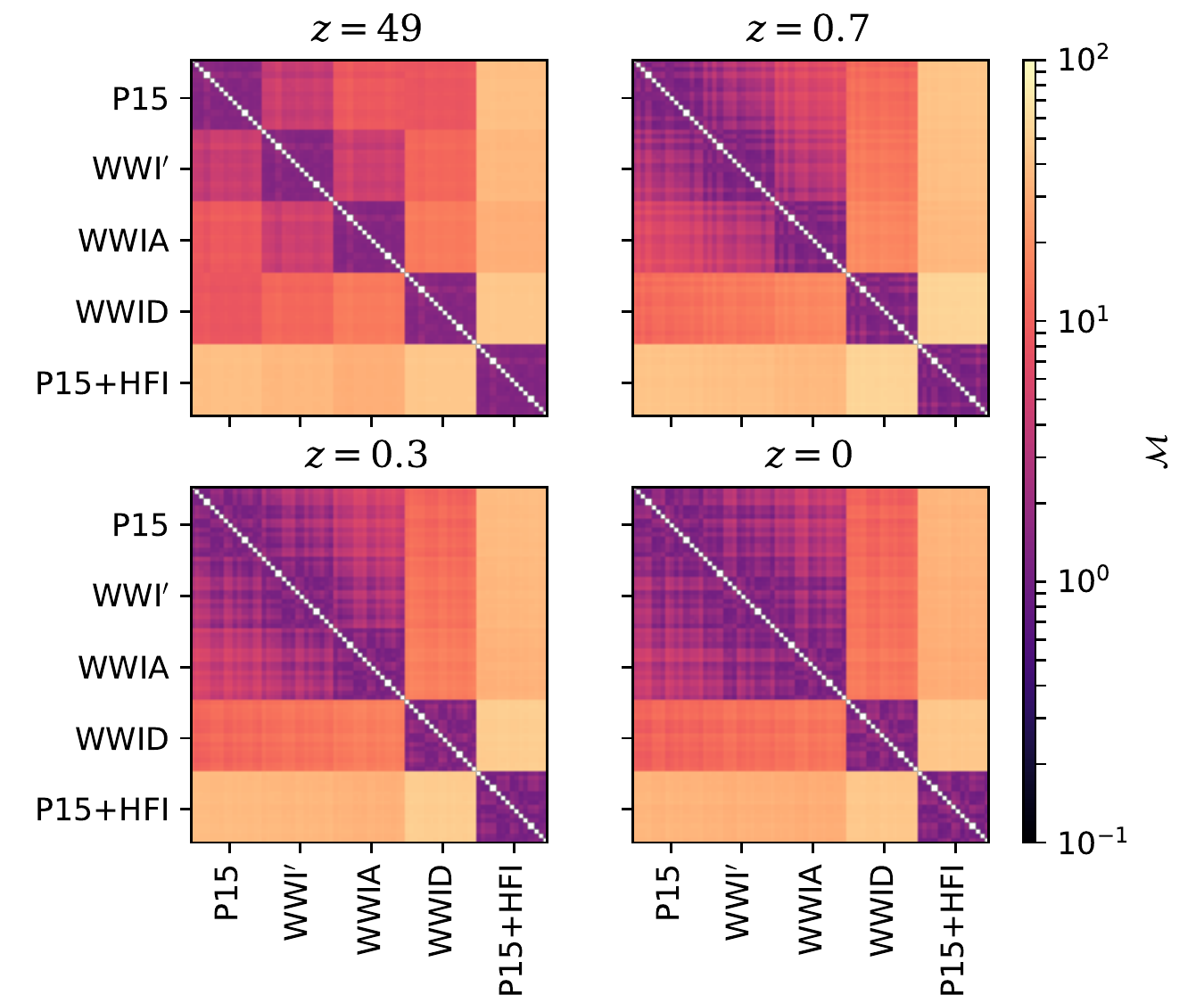}
    \caption{\label{fig:matPk}Separability matrix \tens{M} for the power spectrum in
      Fig.~\ref{fig:Pk}.
    }
  \end{subfigure}
  \begin{subfigure}{.495\textwidth}
    \includegraphics[width=\columnwidth]{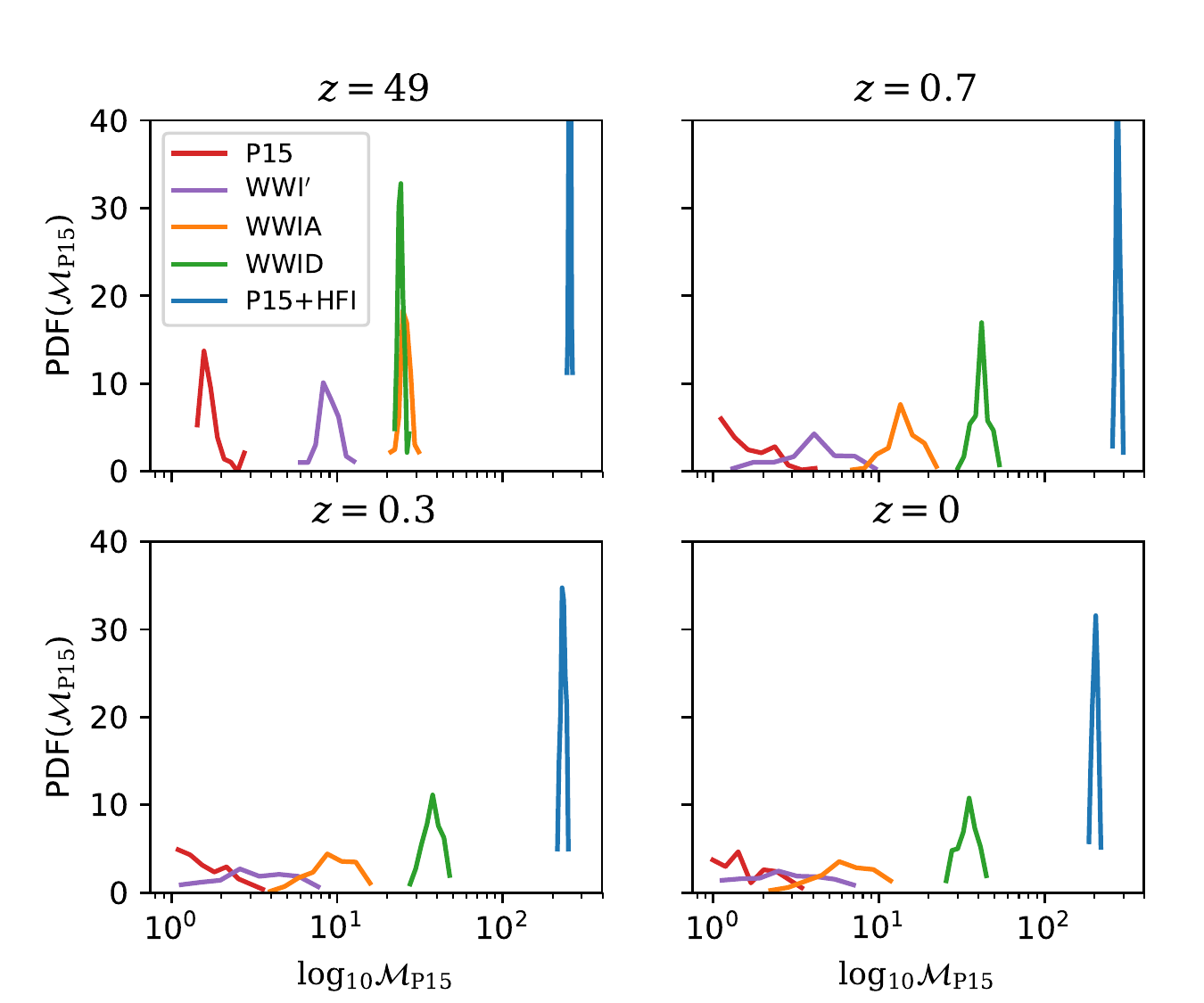}
    \caption{\label{fig:histMPk}%
    Distribution of ${M}_{\mathrm{P15}_i,\text{mod}^n_j}$.}
  \end{subfigure}
  \caption{\label{fig:sepPk}}
\end{figure*}

On smaller  scales ($k\geq \Mpch{0.2}$),  some of the  oscillations in
the WWID PS are still  visible at redshifts of $z\simeq 0.7$). 
As time evolves, these oscillations are smeared out by the mode-mixing
in the non-linear regime, and have essentially disappeared by redshift
0. 
WWID shows some excess of power, consistent with its linear PS and
larger  $\sigma_8$, and  on small  scales ($\geq  \Mpch{0.5}$) can  be
distinguished from P15.  
However, WWIA and WWI$'$ show a very similar PS to P15, and they are not
distinguishable.
This is consistent  with their cosmological parameters  which are very
close to those of P15, and with the small amplitude of the oscillatory
features in their linear PS.

In  order to  quantify the  distance between  any two  simulations, we
  define      the      
  \emph{separability matrix}
  as  the  root  mean  square  of  the
  normalized difference (with respect to P15) between the two simulations:
\begin{equation}
    {M}_{\text{mod}^m_i,\text{mod}^n_j} 
    = \sqrt{\frac 1 {N_k}\sum_k(\Xi_{P_{\text{mod}^m_i}}(k)-\Xi_{P_{\text{mod}^n_j}}(k))^2}, 
\end{equation}
where mod$^n_j$ is the $j$th simulation of model $n$.
\tens{M}  is   thus  constituted  of  $N_\text{models}^2$   blocks  of
$N_\text{real}^2$,
  where $N_\text{models} = 5$ and $N_\text{real} = 15$.
$\tens{M}$  is symmetric  by  construction, and  its component  are
the root mean square of the deviation between two models, which becomes
larger when two models are more distant.
We stress  that the choice of the normalization factor $\sigma_\mathrm{P15}$  
has little impact  on the results, since at a given $k$, all models have similar 
fluctuation.
Fig.  \ref{fig:matPk} shows  the  matrix  $\tens{M}$ corresponding  to
Fig.~\ref{fig:Pk}.
At $z=49$, it  can be easily seen that the  diagonal blocks have lower
values: two simulations from the same model are closer to each other than two simulations from different models. 
Therefore, in the initial conditions, each model can be easily distinguished from the others.
However, as time evolves, some models become confused with others.
For instance, at $z=0$, WWI$'$, WWIA, and P15 show similar values, and
it is difficult to distinguish them.

The matrix \tens{M} contains a huge amount of information, which can be used to distinguish between models in a quantitative way. 
Since our  goal is primarily to  distinguish featured PPS  from the power-law  case, we focus on the reference P15 model, and use \tens{M} for this purpose.
Similar procedure can be applied to any model. 
Fig.~\ref{fig:histMPk}       shows       the      distribution       of the values of
 $M_{\mathrm{P15}_i,\text{mod}^n_j}$, for all realizations $i$ and $j$.
At $z=49$,  the distribution of   $M_{\mathrm{P15}_i,\text{mod}^n_j}$  does not  overlap with those of $M_{\mathrm{P15}_i,\text{mod}^n_j}$ for mod$^n \neq$ P15, confirming  our visual
distinction between the models.
At  lower redshifts,  the  distributions start  to  overlap with  P15:
WWW$'$  from $z=0.7$,  and WWA  from $z=0.3$,  while P15+HFI  and WWID
stay apart.
Therefore,  the distributions  of $M_{\mathrm{P15}_i,\text{mod}^n_j}$  appears
to  be a  useful  metric  to quantify  the  separability of  the
models.
However, these results indicates that the power spectrum is not adequate to distinguish between the models considered in this work as there are overlaps between PDFs at observable redshifts.

\subsection{Halo mass function}

The next natural step after studying the density power spectrum is the
halo mass function  \citep{1974ApJ...187..425P,2007MNRAS.374....2R}. 
As  studied  in  \citet{2013JCAP...03..003H}, features  in  the  power
spectrum are expected  to affect the halo formation rate  and the mass
function.  
However, due to the non-linear  nature of halo formation, passing from
the power spectrum to the halo mass function is not trivial, hence the
need for numerical simulations. 
We identified the haloes in  the simulations with the \code{PFOF} code
\citep{2014A&A...564A..13R},  a massively-parallel  Friends-of-Friends
algorithm \citep{1985ApJ...292..371D}.
Particles  closer than  $b  \,\bar  d$, where  $\bar  d$  is the  mean
particle  separation and  $b$  the  linking length,  set  to 0.2,  are
grouped together.
We  kept all  haloes with  number of  particles $N_\text{p}  \geq 29$,
yielding a minimum mass of \hMsun{1.6e13}, corresponding to groups and 
clusters of galaxies.

\begin{figure}
  \centering
  \includegraphics[width=.5\textwidth]{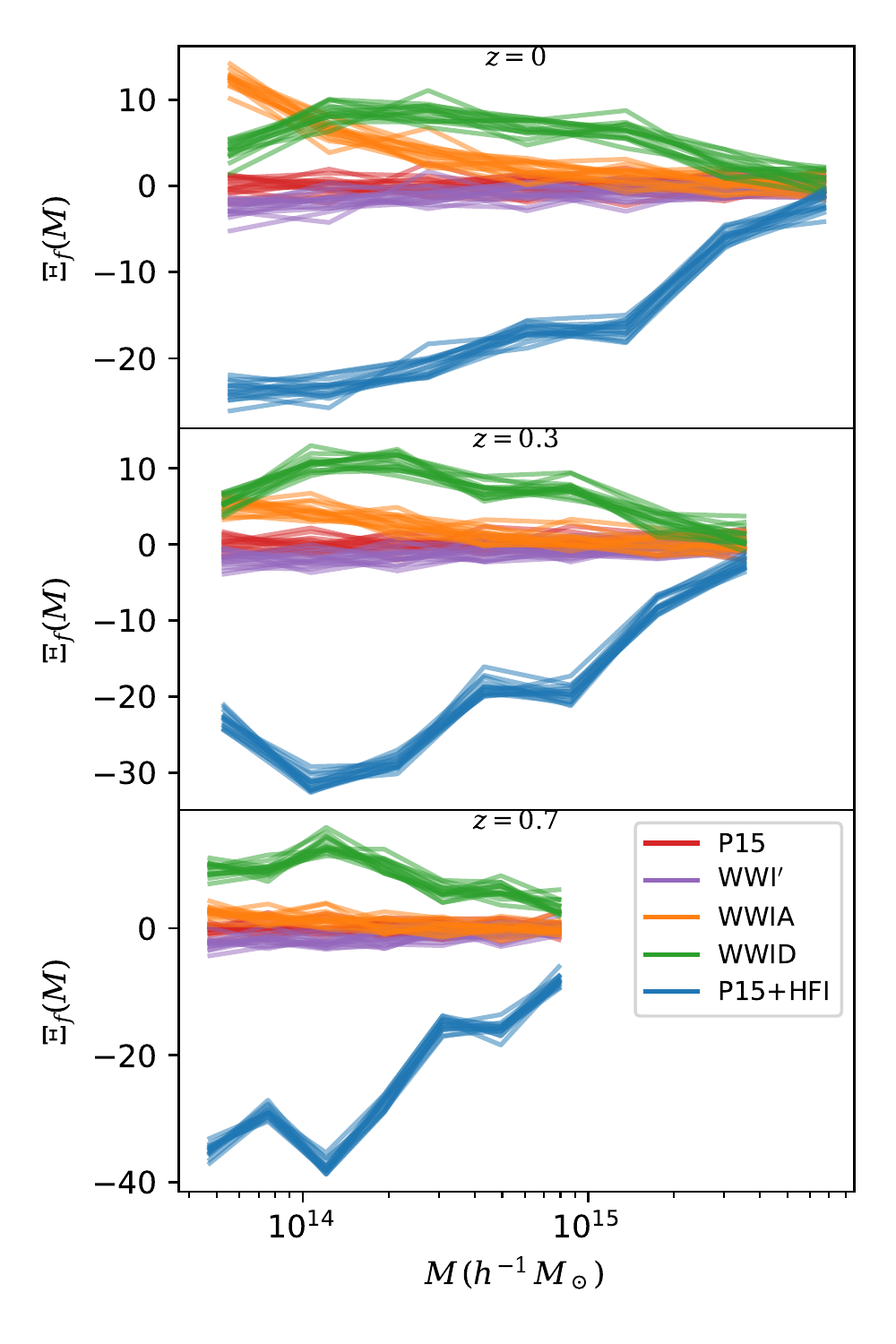}
  \caption{\label{fig:mf}%
    Normalized  difference  of  the halo  mass  function  at
      $z=0,0.3$, and 0.7.    
  }
\end{figure}

Fig.~\ref{fig:mf} shows  the normalized  difference of  the
mass function of simulation $i$ defined as 
\begin{align}
  \label{eq:mf}
\Xi_{f_{i}}(M) & =  \frac{n'_{i}(M)-\mean{n'_\mathrm{P15}(M)}}{\sigma_{P15}(M)},
    \intertext{where}
    n'(M) & =  \deriv n M, 
\end{align}
and  $n(>M)$  is  the cumulative  number  density  of
haloes more massive than $M$, and $\bar \rho$ is the matter density.

Due to its lack of power  on small-scales, P15+HFI yields a lower mass
function, up  to 20\% lower  than P15,  while WWID shows  some excess,
especially at the mass scale of \hMsun{e15}.  
Interestingly,  WWIA can  now  be distinguished  at  $z=0$, while  at
  higher redshift, it is closer to P15.
However, given the small volume at $z=0$, obtaining an accurate mass function may be difficult.
Moreover, WWI$'$ is still indistinguishable from P15.

\begin{figure}
\centering
\includegraphics[width=\columnwidth]{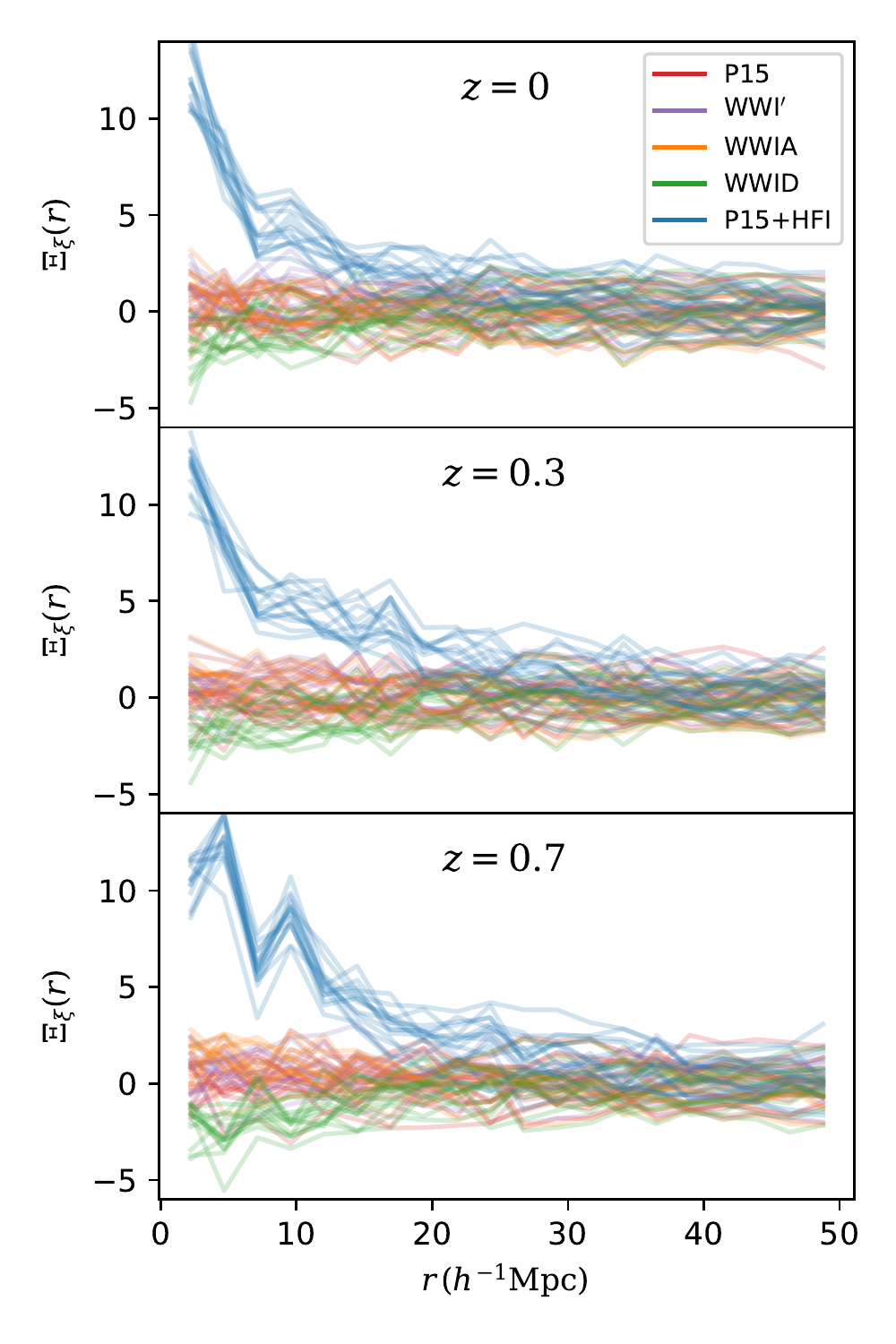}
\caption{\label{fig:xihalo}%
Normalized difference of the  halo correlation function with respect
  to P15. }
\end{figure}

\subsection{Halo two-point correlation function}

Another statistics,  directly comparable to observations,  is the halo
two-point correlation function (2pcf) $\xi$, which measures the excess
of pair clustering at a distance $r$ with respect to the random case.  
We calculated the isotropic two-point correlation $\xi(r)$ function of
FOF       haloes        using       the        \code{kstat}
code.\footnote{\url{https://bitbucket.org/csabiu/kstat}} 
We used the \citet{1993ApJ...412...64L} estimator
\begin{equation}
\xi = \frac{DD-2DR+RR}{RR},
\end{equation}
where DD, DR, and RR are the data-data, data-random, and random-random
pair counts, respectively.

\begin{figure*}
   \includegraphics[width=\textwidth]{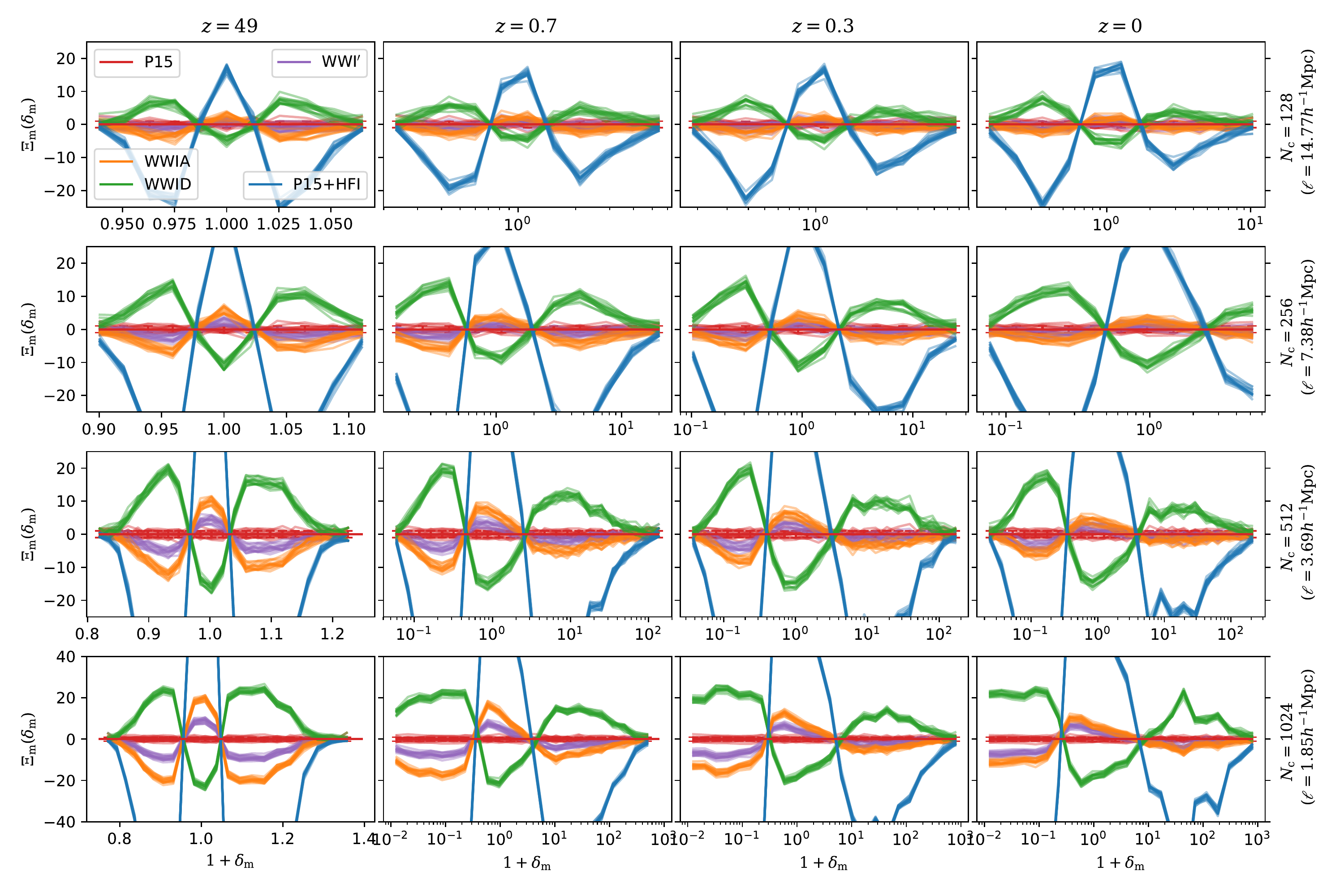}
  \caption{\label{fig:absdiff}
    Normalized  difference  of  the  matter  density
    histogram with the average of the P15 case.
  }
\end{figure*}

Fig.~\ref{fig:xihalo}    shows   the    normalized difference in the 2pcf of 
each simulation with respect to the P15 case, at $z=0$
(top), 0.3 (middle), and 0.7 (bottom).  
Only  P15+HFI,   and  to  a   lesser  extent,  WWID,  can  be
distinguished due to their different  clustering on small
scales, while on large scales, and for other models at all scales, the
fluctuations  are  too large, making
them essentially indistinguishable.

\subsection{Three-dimensional count-in-cell density field}

The previous sections showed the inability of  statistics such as the
power spectrum or the halo mass or two-point correlation function, to
distinguish between WWI$'$ and P15 (however, P15+HFI and WWID
could be distinguished by their  power spectra, and WWIA marginally with
  the mass function at $z=0$).  
Therefore, one needs to find appropriate statistics.
A drawback of the matter power spectrum is that it reduces the
dimensionality of the problem from three to one dimension, losing
information in the process.  
Therefore, we looked at the three-dimensional density field. 

We calculated the probability distribution  function (PDF) of the dark
matter density field binned  in $N_\text{c}^3$ cells, with $N_\text{c}
\in \{128,256,512,1024\}$.  
The  density   field  follows   a  close  to   lognormal  distribution
\citep[e.g.][]{2017ApJ...843...73S}.
The density was calculated using  \code{ComputePk}, with a TSC scheme,
as in \S~\ref{sec:Pk}. 
Fig.~\ref{fig:absdiff}  shows the normalized  difference  of  each simulation  with
  respect to P15.

P15+HFI shows  an excess  probability at $1+\delta  \approx 1$,  and a
lack at  $1+\delta \gg 1$  and $1+\delta \ll  1$ with respect  to P15,
while WWID shows the opposite trend.  
This is  consistent with  the lower amplitude of the  power  spectrum of  P15+HFI, which
yields smaller fluctuations.

Since WWID  and P15+HFI can  be distinguished from the  amplitude of their power spectrum
(see \S~\ref{sec:Pk}), we  will focus on the  remaining models, namely
WWIA and WWI$'$. 
At  the  coarser  level  ($128^3$,  top  row),  even  in  the  initial
conditions, it is not easy to distinguish by eyes the distributions of
the WWIA, WWI$'$, and P15 cases.
As the resolution of the mesh  gets finer (towards lower rows), the PDFs
show more and more difference.
For  intermediate  meshes  (256  and   512  cases),  the  PDF  can  be
distinguished in  the initial conditions  and up to redshift  0.7, and
for the $N_\text{c} = 1024^3$ case, all models have their PDF clearly
separated from the other models. 

\begin{figure*}
\centering
\includegraphics[width=\textwidth]{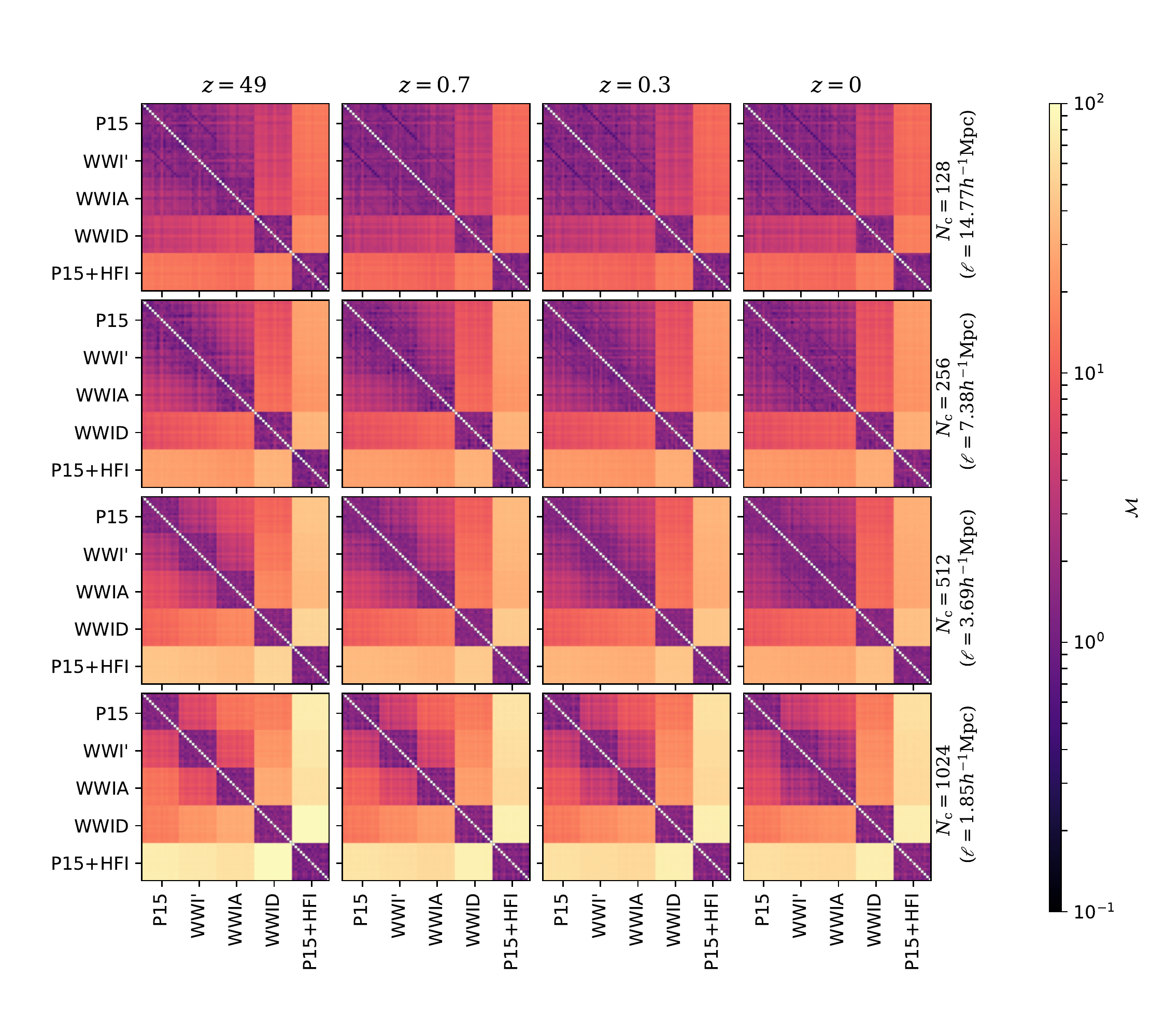}
\caption{\label{fig:mat_dens}Separability matrix of the matter density}
\end{figure*}

\begin{figure*}
  \centering
  \includegraphics[width=\textwidth]{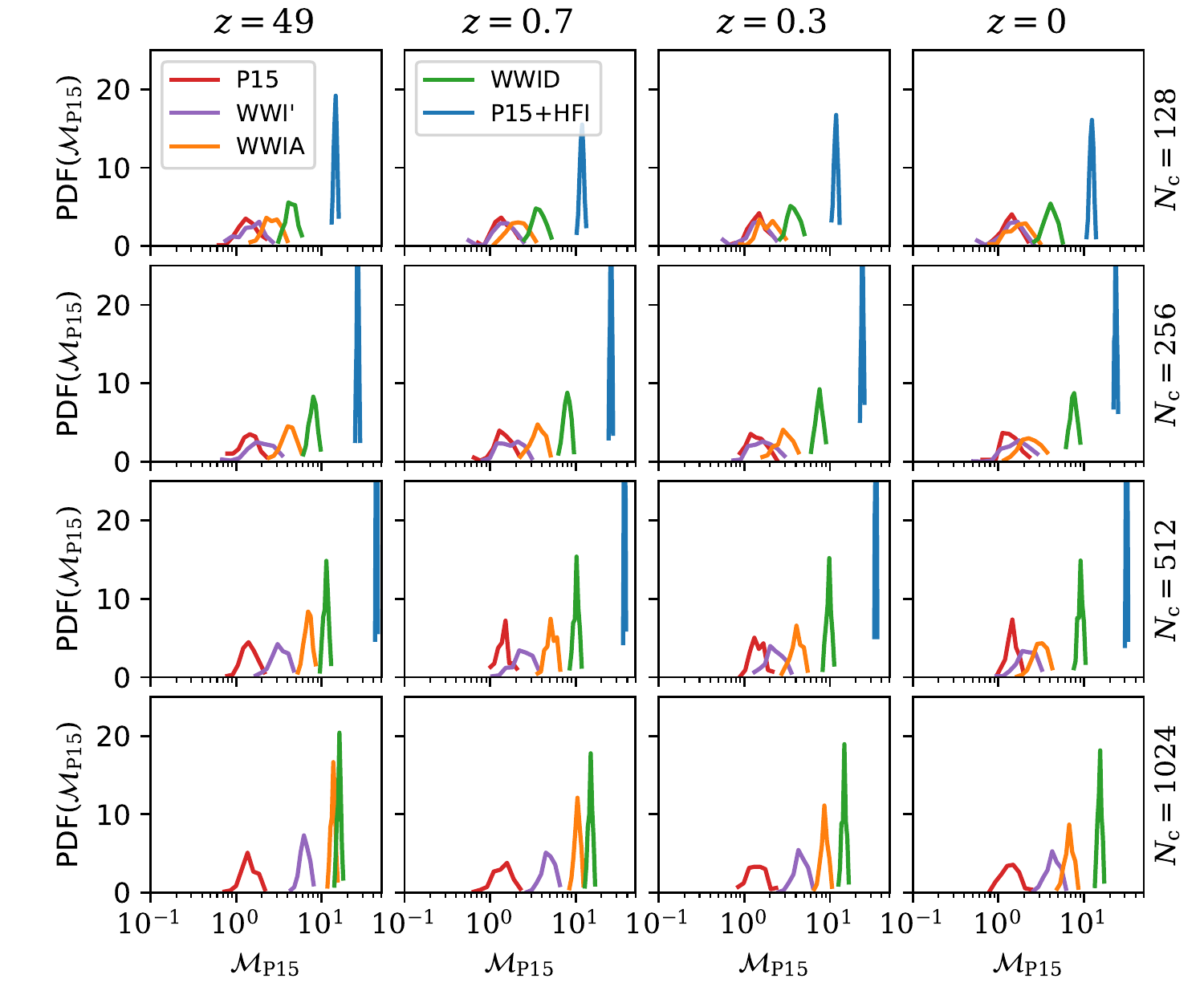}
  \caption{\label{fig:histmat_dens}Distribution of $M_{\mathrm{P15}_i,\text{mod}^n_j}$ for the matter density shown in Figs.~\ref{fig:absdiff} and \ref{fig:mat_dens}.}
\end{figure*}

To quantify the  separability of the models, we  show the
separability matrix \tens{M} of the matter density distribution
in Fig.~\ref{fig:mat_dens}.
Clearly, for low-resolution cases ($N_\text{c}  = 128$ or 256), only in
the  initial density  field we can distinguish  between the  models,
while  for   high  enough  resolution   (512 or 1024),  all  models   can  be
distinguished.
This is  further supported by  the distribution  of $M_{\mathrm{P15}_i,\text{mod}^n_j}$,
the first column of the matrix, shown in Fig.~\ref{fig:histmat_dens}.
While for the previous estimators, the distribution of  $M_{\mathrm{P15}_i,\text{P15}_j}$ and $M_{\mathrm{P15}_i,\text{WWI}'_j}$ were overlapping, they are now well separated at $z = 0.7$.
Therefore, the three-dimensional matter density field is able to distinguish between all our models presented here, at least at $z=0.7$.

\begin{figure*}
\includegraphics[width=\textwidth]{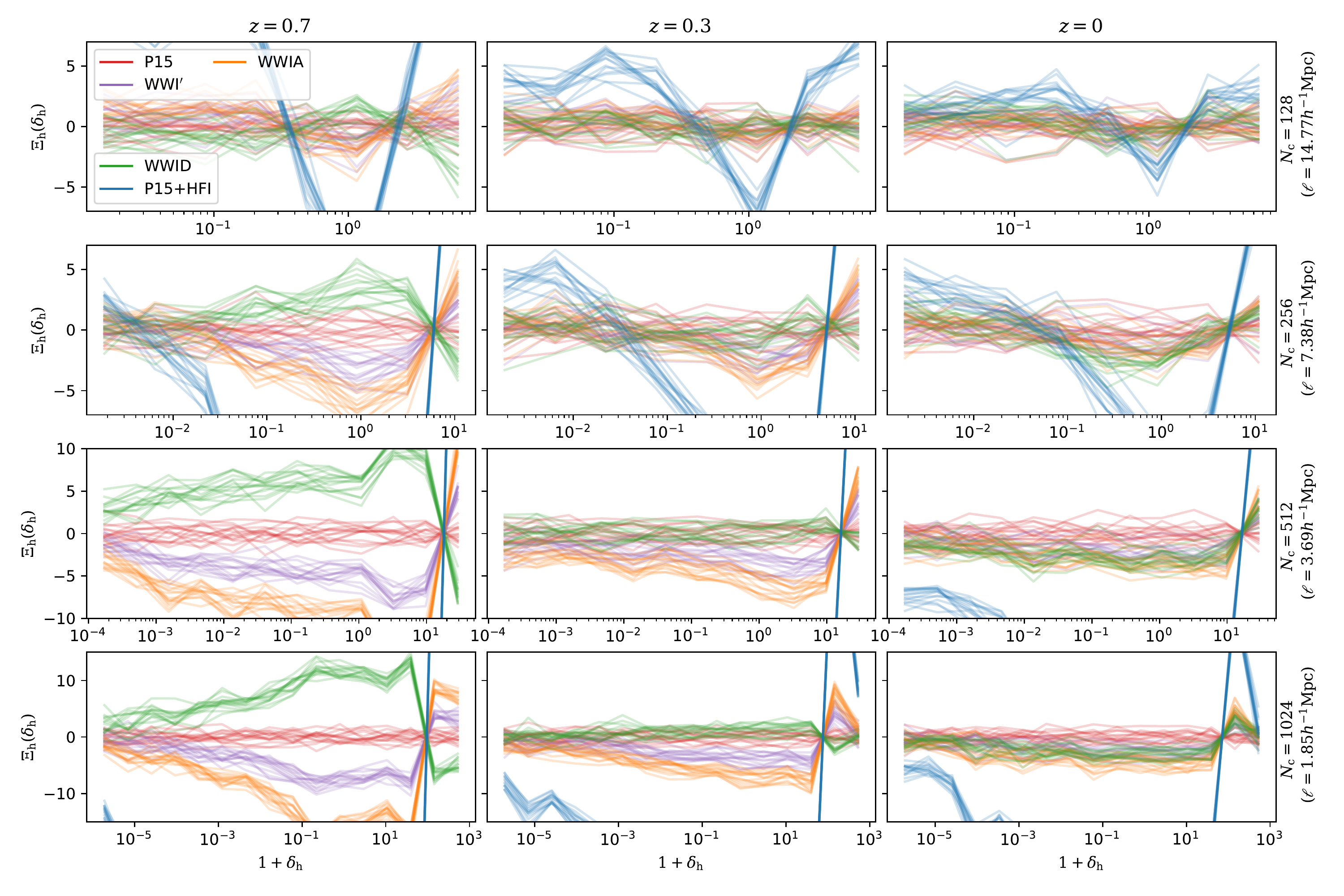}
\caption{\label{fig:dhalo}
Normalized difference of  the PDFs of  the mass-weighted halo  density with
respect to  the P15 case  for $N_\text{c} =  128$, 256, 512,  and 1024
(from top to bottom). 
}
\end{figure*}

\begin{figure*}
  \centering
  \includegraphics[width=\textwidth]{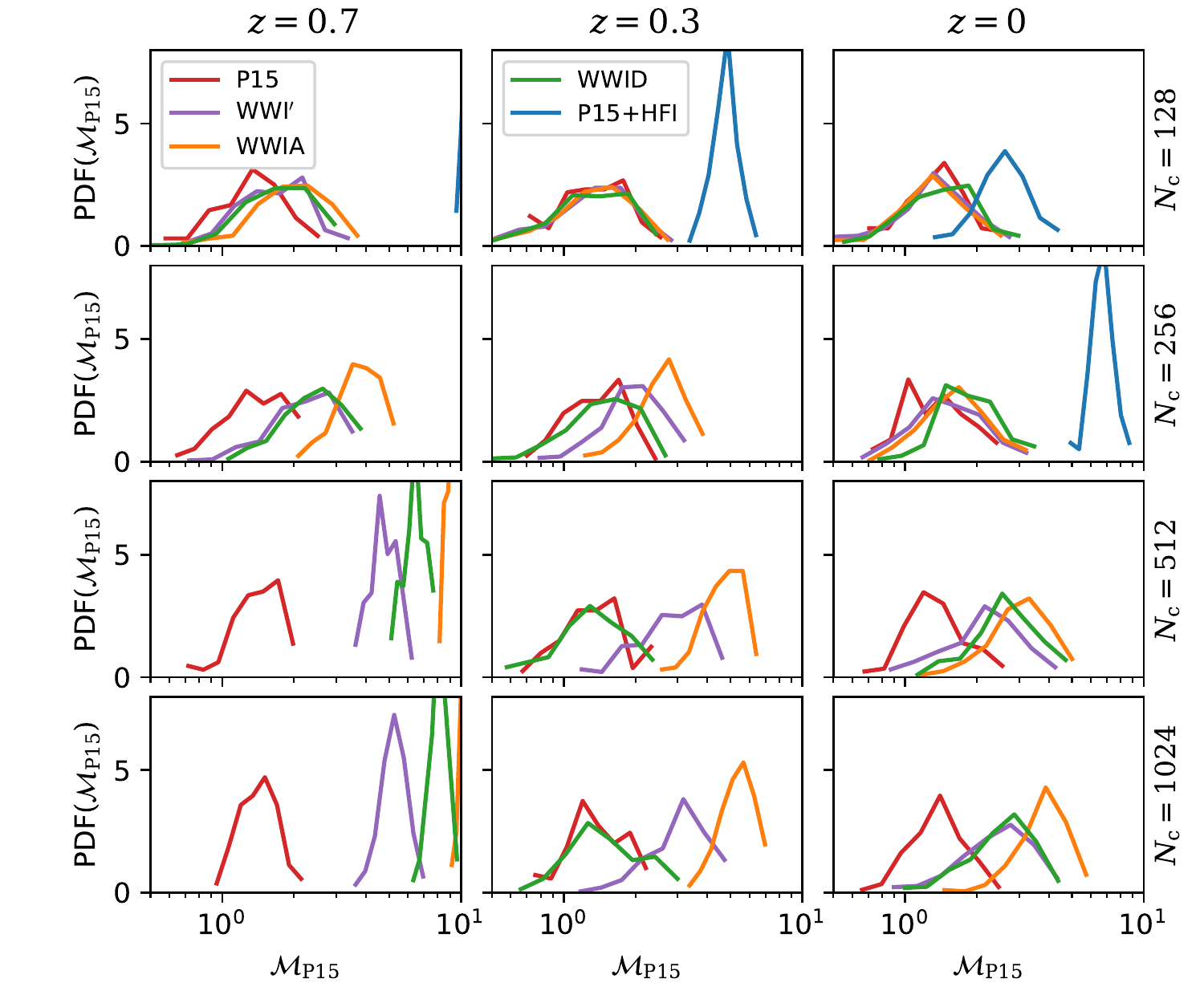}
  \caption{\label{fig:histM_dhalo}Distribution                      of
    $M_{\mathrm{P15}_i,\text{mod}^n_j}$ for the halo density distribution.}
\end{figure*}

The normalized difference in the matter  density count-in-cell gives us important
hint as to how the EU affects the LSS.  
However, matter is not directly  observable, and we have to understand
how biased tracers such as galaxies are affected.  
For that purpose, we applied the same technique to FoF haloes.

Fig.~\ref{fig:dhalo}  shows the normalized difference in  the PDFs  of the
mass-weighted  halo density  with respect  to the  average of  the P15
case.  
The distribution of $M_{\text{P15}_i,\text{mod}^n_j}$ is shown in Fig.~\ref{fig:histM_dhalo}.
We   used   the   mass-weighted     rather   than   
number-weighted halo   density, since the former shows a tighter correlation with the
matter density field \citep{2012ApJ...753...11J, 2018MNRAS.473.5098U}.  
In the  $N_\text{c}=128$ case, the  statistics are very noisy,  due to
the  small  number   of  pixels,  which  only   marginally  allows us  to
distinguish P15+HFI from the rest: all models are visually indistinguishable in Fig.~\ref{fig:dhalo}, which is confirmed by the overlap of the distributions of $M_{\text{P15}_i,\text{mod}^n_j}$ in Fig.~\ref{fig:histM_dhalo}.  
When  going to  smaller  pixels, it  becomes  possible to  distinguish
between the models at $z=0.7$ for $N_\text{c}\geq 512$. 
In  Fig.~\ref{fig:histM_dhalo}, it can clearly be seen that the distribution of $M_{\text{P15}_i,\text{P15}_j}$ (red) is well separated from the other models $M_{\text{P15}_i,\text{mod}^n_j}$.

\section{Discussion and summary}

\label{sec:ccl}

In order  to distinguish  five different  degenerate (with  respect to
the CMB data) forms of the PPS (Planck 2015 best-fit, Planck 2015+HFI,
and  the  best-fitting wiggly-whipped  inflation  models  WWIA, WWID,  and
WWI$'$),  using the  large-scale  structures,  we ran  a  series of  15
DESI-like $N$-body simulations.  

We measured different  statistics, such as the  density power spectrum,
halo  mass   function,  halo   two-point  correlation   function,  and
count-in-cell density (matter and mass-weighted halo) in order to assess 
their power to distinguish between these models. 
We then introduced a separability matrix to differentiate quantitatively 
between different models using our measured statistics from the simulations.

While the PS  and HMF can distinguish between  certain models (P15+HFI
and WWID), the remaining three models are indistinguishable from P15. 

Instead of  reducing the dimensionality  of the problem, we  work with
the  three-dimensional density  field,  taking advantage  of the  huge
statistics from the large simulation volume. 
For  a   DESI-like  survey,   when  using  enough   pixels  ($1024^3$,
corresponding  to  a cube  of  size  $l=\hMpc{1.85}$), the  difference
between the  PDFs of  the matter density  field allows us  to distinguish
unambiguously between the models. 
Moreover, even  when moving  from the matter-case  to the  biased halo
case, the models can be distinguished.
At $z=0.7$,  the distributions of  the mass-weighted halo  density are
still   distinguishable,  but   at  $z=0.3$,   only  P15+HFI   can  be
distinguished.  
This is the range that will be probed by surveys like DESI, Euclid, or 
LSST.

It should be noted that other studies have been done on  using  the LSS  to  constrain
  features                 in                  the                 PPS
  \citep[e.g.,][]{2016JCAP...10..041B, 2017arXiv171002570P, 2018JCAP...01..051A}. 
  However, they focus on the linear  regime, or use the extended Press
  \& Schechter formalism to study the nonlinear regime.
  These works also rely mostly on some parametrization  of the bias in order to go from matter to galaxy power spectrum. In this  work, we make a step further, considering a  finite volume (corresponding to the DESI volume), and ran $N$-body simulations for a more realistic study without incorporating any parametrization. 

At the end we should also mention that there  is a very  little difference
between the observables of the WWI  models and the reference power law
model.
While,  in  this  work,  we  did not  consider  varying  the  background
cosmological  parameters, in  reality we  might  have to  face a  more
difficult  problem considering  cosmographic degeneracies  between the
background parameters and the form of the PPS.
This  seems  to  be  an  area which  requires  substantial  amount  of
analysis,  computation and  implementation of  appropriate statistical
approaches in  order to  be prepared  for the  next generation  of the
cosmological observations to use the  information to get closer to the
actual model of the EU.

\section*{Acknowledgements}
We thank Eric Linder, Changbom Park, and Stephen Appleby for their 
stimulating comments.  
This work  was supported by the  Supercomputing Center/Korea Institute
of Science  and Technology  Information with  supercomputing resources
including technical support (KSC-2015-C1-014 and KSC-2016-C2-0035). 
The  post-processing  was  performed  by using  the  high  performance
computing cluster  Polaris at  the Korea  Astronomy and  Space Science
Institute.  
A.S.  would  like  to  acknowledge  the  support  of  the  
National  Research  Foundation  of  Korea (NRF-2016R1C1B2016478).
A.A.S. was partially supported by the Scientific Programme 28
(sub-programme II) of the Presidium of the Russian Academy of Sciences.



\bibliographystyle{mnras}
\bibliography{biblio} 

\begin{thebibliography}{}
\makeatletter
\relax
\def\mn@urlcharsother{\let\do\@makeother \do\$\do\&\do\#\do\^\do\_\do\%\do\~}
\def\mn@doi{\begingroup\mn@urlcharsother \@ifnextchar [ {\mn@doi@}
  {\mn@doi@[]}}
\def\mn@doi@[#1]#2{\def\@tempa{#1}\ifx\@tempa\@empty \href
  {http://dx.doi.org/#2} {doi:#2}\else \href {http://dx.doi.org/#2} {#1}\fi
  \endgroup}
\def\mn@eprint#1#2{\mn@eprint@#1:#2::\@nil}
\def\mn@eprint@arXiv#1{\href {http://arxiv.org/abs/#1} {{\tt arXiv:#1}}}
\def\mn@eprint@dblp#1{\href {http://dblp.uni-trier.de/rec/bibtex/#1.xml}
  {dblp:#1}}
\def\mn@eprint@#1:#2:#3:#4\@nil{\def\@tempa {#1}\def\@tempb {#2}\def\@tempc
  {#3}\ifx \@tempc \@empty \let \@tempc \@tempb \let \@tempb \@tempa \fi \ifx
  \@tempb \@empty \def\@tempb {arXiv}\fi \@ifundefined
  {mn@eprint@\@tempb}{\@tempb:\@tempc}{\expandafter \expandafter \csname
  mn@eprint@\@tempb\endcsname \expandafter{\@tempc}}}

\bibitem[\protect\citeauthoryear{{Adams}, {Cresswell}  \& {Easther}}{{Adams}
  et~al.}{2001}]{2001PhRvD..64l3514A}
{Adams} J.,  {Cresswell} B.,   {Easther} R.,  2001, \mn@doi [\prd]
  {10.1103/PhysRevD.64.123514}, \href
  {http://adsabs.harvard.edu/abs/2001PhRvD..64l3514A} {64, 123514}

\bibitem[\protect\citeauthoryear{{Alimi} et~al.,}{{Alimi}
  et~al.}{2012}]{2012arXiv1206.2838A}
{Alimi} J.-M.,  et~al., 2012, preprint, \href
  {http://adsabs.harvard.edu/abs/2012arXiv1206.2838A} {} (\mn@eprint {arXiv}
  {1206.2838})

\bibitem[\protect\citeauthoryear{{Ansari Fard} \& {Baghram}}{{Ansari Fard} \&
  {Baghram}}{2018}]{2018JCAP...01..051A}
{Ansari Fard} M.,  {Baghram} S.,  2018, \mn@doi [\jcap]
  {10.1088/1475-7516/2018/01/051}, \href
  {http://adsabs.harvard.edu/abs/2018JCAP...01..051A} {1, 051}

\bibitem[\protect\citeauthoryear{{Baldi}}{{Baldi}}{2012}]{2012MNRAS.422.1028B}
{Baldi} M.,  2012, \mn@doi [\mnras] {10.1111/j.1365-2966.2012.20675.x}, \href
  {http://adsabs.harvard.edu/abs/2012MNRAS.422.1028B} {422, 1028}

\bibitem[\protect\citeauthoryear{{Baldi}, {Villaescusa-Navarro}, {Viel},
  {Puchwein}, {Springel}  \& {Moscardini}}{{Baldi}
  et~al.}{2014}]{2014MNRAS.440...75B}
{Baldi} M.,  {Villaescusa-Navarro} F.,  {Viel} M.,  {Puchwein} E.,  {Springel}
  V.,   {Moscardini} L.,  2014, \mn@doi [\mnras] {10.1093/mnras/stu259}, \href
  {http://adsabs.harvard.edu/abs/2014MNRAS.440...75B} {440, 75}

\bibitem[\protect\citeauthoryear{{Ballardini}, {Finelli}, {Fedeli}  \&
  {Moscardini}}{{Ballardini} et~al.}{2016}]{2016JCAP...10..041B}
{Ballardini} M.,  {Finelli} F.,  {Fedeli} C.,   {Moscardini} L.,  2016, \mn@doi
  [\jcap] {10.1088/1475-7516/2016/10/041}, \href
  {http://adsabs.harvard.edu/abs/2016JCAP...10..041B} {10, 041}

\bibitem[\protect\citeauthoryear{{Brax}, {Davis}, {Li}, {Winther}  \&
  {Zhao}}{{Brax} et~al.}{2012}]{2012JCAP...10..002B}
{Brax} P.,  {Davis} A.-C.,  {Li} B.,  {Winther} H.~A.,   {Zhao} G.-B.,  2012,
  \mn@doi [\jcap] {10.1088/1475-7516/2012/10/002}, \href
  {http://adsabs.harvard.edu/abs/2012JCAP...10..002B} {10, 002}

\bibitem[\protect\citeauthoryear{{Brax}, {Davis}, {Li}, {Winther}  \&
  {Zhao}}{{Brax} et~al.}{2013}]{2013JCAP...04..029B}
{Brax} P.,  {Davis} A.-C.,  {Li} B.,  {Winther} H.~A.,   {Zhao} G.-B.,  2013,
  \mn@doi [\jcap] {10.1088/1475-7516/2013/04/029}, \href
  {http://adsabs.harvard.edu/abs/2013JCAP...04..029B} {4, 029}

\bibitem[\protect\citeauthoryear{{Crocce}, {Pueblas}  \&
  {Scoccimarro}}{{Crocce} et~al.}{2006}]{2006MNRAS.373..369C}
{Crocce} M.,  {Pueblas} S.,   {Scoccimarro} R.,  2006, \mn@doi [\mnras]
  {10.1111/j.1365-2966.2006.11040.x}, \href
  {http://adsabs.harvard.edu/abs/2006MNRAS.373..369C} {373, 369}

\bibitem[\protect\citeauthoryear{{DESI Collaboration} et~al.,}{{DESI
  Collaboration} et~al.}{2016a}]{2016arXiv161100036D}
{DESI Collaboration} et~al., 2016a, preprint, \href
  {http://adsabs.harvard.edu/abs/2016arXiv161100036D} {} (\mn@eprint {arXiv}
  {1611.00036})

\bibitem[\protect\citeauthoryear{{DESI Collaboration} et~al.,}{{DESI
  Collaboration} et~al.}{2016b}]{2016arXiv161100037D}
{DESI Collaboration} et~al., 2016b, preprint, \href
  {http://adsabs.harvard.edu/abs/2016arXiv161100037D} {} (\mn@eprint {arXiv}
  {1611.00037})

\bibitem[\protect\citeauthoryear{{Davis}, {Efstathiou}, {Frenk}  \&
  {White}}{{Davis} et~al.}{1985}]{1985ApJ...292..371D}
{Davis} M.,  {Efstathiou} G.,  {Frenk} C.~S.,   {White} S.~D.~M.,  1985,
  \mn@doi [\apj] {10.1086/163168}, \href
  {http://adsabs.harvard.edu/abs/1985ApJ...292..371D} {292, 371}

\bibitem[\protect\citeauthoryear{{Guth}}{{Guth}}{1981}]{1981PhRvD..23..347G}
{Guth} A.~H.,  1981, \mn@doi [\prd] {10.1103/PhysRevD.23.347}, \href
  {http://adsabs.harvard.edu/abs/1981PhRvD..23..347G} {23, 347}

\bibitem[\protect\citeauthoryear{{Hazra}}{{Hazra}}{2013}]{2013JCAP...03..003H}
{Hazra} D.~K.,  2013, \mn@doi [\jcap] {10.1088/1475-7516/2013/03/003}, \href
  {http://adsabs.harvard.edu/abs/2013JCAP...03..003H} {3, 003}

\bibitem[\protect\citeauthoryear{{Hazra}, {Shafieloo}, {Smoot}  \&
  {Starobinsky}}{{Hazra} et~al.}{2014a}]{2014JCAP...08..048H}
{Hazra} D.~K.,  {Shafieloo} A.,  {Smoot} G.~F.,   {Starobinsky} A.~A.,  2014a,
  \mn@doi [\jcap] {10.1088/1475-7516/2014/08/048}, \href
  {http://adsabs.harvard.edu/abs/2014JCAP...08..048H} {8, 048}

\bibitem[\protect\citeauthoryear{{Hazra}, {Shafieloo}, {Smoot}  \&
  {Starobinsky}}{{Hazra} et~al.}{2014b}]{2014PhRvL.113g1301H}
{Hazra} D.~K.,  {Shafieloo} A.,  {Smoot} G.~F.,   {Starobinsky} A.~A.,  2014b,
  \mn@doi [Physical Review Letters] {10.1103/PhysRevLett.113.071301}, \href
  {http://adsabs.harvard.edu/abs/2014PhRvL.113g1301H} {113, 071301}

\bibitem[\protect\citeauthoryear{{Hazra}, {Shafieloo}, {Smoot}  \&
  {Starobinsky}}{{Hazra} et~al.}{2016}]{2016JCAP...09..009H}
{Hazra} D.~K.,  {Shafieloo} A.,  {Smoot} G.~F.,   {Starobinsky} A.~A.,  2016,
  \mn@doi [\jcap] {10.1088/1475-7516/2016/09/009}, \href
  {http://adsabs.harvard.edu/abs/2016JCAP...09..009H} {9, 009}

\bibitem[\protect\citeauthoryear{{Hazra}, {Paoletti}, {Ballardini}, {Finelli},
  {Shafieloo}, {Smoot}  \& {Starobinsky}}{{Hazra}
  et~al.}{2018}]{2018JCAP...02..017H}
{Hazra} D.~K.,  {Paoletti} D.,  {Ballardini} M.,  {Finelli} F.,  {Shafieloo}
  A.,  {Smoot} G.~F.,   {Starobinsky} A.~A.,  2018, \mn@doi [\jcap]
  {10.1088/1475-7516/2018/02/017}, \href
  {http://adsabs.harvard.edu/abs/2018JCAP...02..017H} {2, 017}

\bibitem[\protect\citeauthoryear{{Hellwing}, {Frenk}, {Cautun}, {Bose},
  {Helly}, {Jenkins}, {Sawala}  \& {Cytowski}}{{Hellwing}
  et~al.}{2016}]{2016MNRAS.457.3492H}
{Hellwing} W.~A.,  {Frenk} C.~S.,  {Cautun} M.,  {Bose} S.,  {Helly} J.,
  {Jenkins} A.,  {Sawala} T.,   {Cytowski} M.,  2016, \mn@doi [\mnras]
  {10.1093/mnras/stw214}, \href
  {http://adsabs.harvard.edu/abs/2016MNRAS.457.3492H} {457, 3492}

\bibitem[\protect\citeauthoryear{{Hockney} \& {Eastwood}}{{Hockney} \&
  {Eastwood}}{1988}]{1988csup.book.....H}
{Hockney} R.~W.,  {Eastwood} J.~W.,  1988, {Computer simulation using
  particles}.
Bristol

\bibitem[\protect\citeauthoryear{{Horiguchi}, {Ichiki}  \&
  {Yokoyama}}{{Horiguchi} et~al.}{2017}]{2017PTEP.2017i3E01H}
{Horiguchi} K.,  {Ichiki} K.,   {Yokoyama} J.,  2017, \mn@doi [Progress of
  Theoretical and Experimental Physics] {10.1093/ptep/ptx121}, \href
  {http://adsabs.harvard.edu/abs/2017PTEP.2017i3E01H} {2017, 093E01}

\bibitem[\protect\citeauthoryear{{Ivezic} et~al.,}{{Ivezic}
  et~al.}{2008}]{2008arXiv0805.2366I}
{Ivezic} Z.,  et~al., 2008, preprint, \href
  {http://adsabs.harvard.edu/abs/2008arXiv0805.2366I} {} (\mn@eprint {arXiv}
  {0805.2366})

\bibitem[\protect\citeauthoryear{{Jee}, {Park}, {Kim}, {Choi}  \& {Kim}}{{Jee}
  et~al.}{2012}]{2012ApJ...753...11J}
{Jee} I.,  {Park} C.,  {Kim} J.,  {Choi} Y.-Y.,   {Kim} S.~S.,  2012, \mn@doi
  [\apj] {10.1088/0004-637X/753/1/11}, \href
  {http://adsabs.harvard.edu/abs/2012ApJ...753...11J} {753, 11}

\bibitem[\protect\citeauthoryear{{Joy}, {Sahni}  \& {Starobinsky}}{{Joy}
  et~al.}{2008}]{2008PhRvD..77b3514J}
{Joy} M.,  {Sahni} V.,   {Starobinsky} A.~A.,  2008, \mn@doi [\prd]
  {10.1103/PhysRevD.77.023514}, \href
  {http://adsabs.harvard.edu/abs/2008PhRvD..77b3514J} {77, 023514}

\bibitem[\protect\citeauthoryear{{Joy}, {Shafieloo}, {Sahni}  \&
  {Starobinsky}}{{Joy} et~al.}{2009}]{2009JCAP...06..028J}
{Joy} M.,  {Shafieloo} A.,  {Sahni} V.,   {Starobinsky} A.~A.,  2009, \mn@doi
  [\jcap] {10.1088/1475-7516/2009/06/028}, \href
  {http://adsabs.harvard.edu/abs/2009JCAP...06..028J} {6, 028}

\bibitem[\protect\citeauthoryear{{L'Huillier}}{{L'Huillier}}{2014}]{2014ascl.soft03015L}
{L'Huillier} B.,  2014, {computePk: Power spectrum computation} (\mn@eprint
  {ascl} {1403.015})

\bibitem[\protect\citeauthoryear{{L'Huillier}, {Park}  \& {Kim}}{{L'Huillier}
  et~al.}{2014}]{2014NewA...30...79L}
{L'Huillier} B.,  {Park} C.,   {Kim} J.,  2014, \mn@doi [\na]
  {10.1016/j.newast.2014.01.007}, \href
  {http://adsabs.harvard.edu/abs/2014NewA...30...79L} {30, 79}

\bibitem[\protect\citeauthoryear{{L'Huillier}, {Winther}, {Mota}, {Park}  \&
  {Kim}}{{L'Huillier} et~al.}{2017}]{2017MNRAS.468.3174L}
{L'Huillier} B.,  {Winther} H.~A.,  {Mota} D.~F.,  {Park} C.,   {Kim} J.,
  2017, \mn@doi [\mnras] {10.1093/mnras/stx700}, \href
  {http://adsabs.harvard.edu/abs/2017MNRAS.468.3174L} {468, 3174}

\bibitem[\protect\citeauthoryear{{Landy} \& {Szalay}}{{Landy} \&
  {Szalay}}{1993}]{1993ApJ...412...64L}
{Landy} S.~D.,  {Szalay} A.~S.,  1993, \mn@doi [\apj] {10.1086/172900}, \href
  {http://adsabs.harvard.edu/abs/1993ApJ...412...64L} {412, 64}

\bibitem[\protect\citeauthoryear{{Laureijs} et~al.,}{{Laureijs}
  et~al.}{2011}]{2011arXiv1110.3193L}
{Laureijs} R.,  et~al., 2011, preprint, \href
  {http://adsabs.harvard.edu/abs/2011arXiv1110.3193L} {} (\mn@eprint {arXiv}
  {1110.3193})

\bibitem[\protect\citeauthoryear{{Lewis}, {Challinor}  \& {Lasenby}}{{Lewis}
  et~al.}{2000}]{2000ApJ...538..473L}
{Lewis} A.,  {Challinor} A.,   {Lasenby} A.,  2000, \mn@doi [\apj]
  {10.1086/309179}, \href {http://adsabs.harvard.edu/abs/2000ApJ...538..473L}
  {538, 473}

\bibitem[\protect\citeauthoryear{{Linde}}{{Linde}}{1982}]{1982PhLB..108..389L}
{Linde} A.~D.,  1982, \mn@doi [Physics Letters B]
  {10.1016/0370-2693(82)91219-9}, \href
  {http://adsabs.harvard.edu/abs/1982PhLB..108..389L} {108, 389}

\bibitem[\protect\citeauthoryear{{Llinares} \& {Mota}}{{Llinares} \&
  {Mota}}{2014}]{2014PhRvD..89h4023L}
{Llinares} C.,  {Mota} D.~F.,  2014, \mn@doi [\prd]
  {10.1103/PhysRevD.89.084023}, \href
  {http://adsabs.harvard.edu/abs/2014PhRvD..89h4023L} {89, 084023}

\bibitem[\protect\citeauthoryear{{Mukhanov} \& {Chibisov}}{{Mukhanov} \&
  {Chibisov}}{1981}]{1981JETPL..33..532M}
{Mukhanov} V.~F.,  {Chibisov} G.~V.,  1981, Soviet Journal of Experimental and
  Theoretical Physics Letters, \href
  {http://adsabs.harvard.edu/abs/1981JETPL..33..532M} {33, 532}

\bibitem[\protect\citeauthoryear{{Palma}, {Sapone}  \& {Sypsas}}{{Palma}
  et~al.}{2017}]{2017arXiv171002570P}
{Palma} G.~A.,  {Sapone} D.,   {Sypsas} S.,  2017, preprint, \href
  {http://adsabs.harvard.edu/abs/2017arXiv171002570P} {} (\mn@eprint {arXiv}
  {1710.02570})

\bibitem[\protect\citeauthoryear{{Planck Collaboration XIII}}{{Planck
  Collaboration XIII}}{2016}]{2016A&A...594A..13P}
{Planck Collaboration XIII} 2016, \mn@doi [\aap] {10.1051/0004-6361/201525830},
  \href {http://adsabs.harvard.edu/abs/2016A%26A...594A..13P} {594, A13}

\bibitem[\protect\citeauthoryear{{Planck Collaboration XX}}{{Planck
  Collaboration XX}}{2016}]{2016A&A...594A..20P}
{Planck Collaboration XX} 2016, \mn@doi [\aap] {10.1051/0004-6361/201525898},
  \href {http://adsabs.harvard.edu/abs/2016A%26A...594A..20P} {594, A20}

\bibitem[\protect\citeauthoryear{{Press} \& {Schechter}}{{Press} \&
  {Schechter}}{1974}]{1974ApJ...187..425P}
{Press} W.~H.,  {Schechter} P.,  1974, \mn@doi [\apj] {10.1086/152650}, \href
  {http://adsabs.harvard.edu/abs/1974ApJ...187..425P} {187, 425}

\bibitem[\protect\citeauthoryear{{Reed}, {Bower}, {Frenk}, {Jenkins}  \&
  {Theuns}}{{Reed} et~al.}{2007}]{2007MNRAS.374....2R}
{Reed} D.~S.,  {Bower} R.,  {Frenk} C.~S.,  {Jenkins} A.,   {Theuns} T.,  2007,
  \mn@doi [\mnras] {10.1111/j.1365-2966.2006.11204.x}, \href
  {http://adsabs.harvard.edu/abs/2007MNRAS.374....2R} {374, 2}

\bibitem[\protect\citeauthoryear{{Roy}, {Bouillot}  \& {Rasera}}{{Roy}
  et~al.}{2014}]{2014A&A...564A..13R}
{Roy} F.,  {Bouillot} V.~R.,   {Rasera} Y.,  2014, \mn@doi [\aap]
  {10.1051/0004-6361/201322555}, \href
  {http://adsabs.harvard.edu/abs/2014A%26A...564A..13R} {564, A13}

\bibitem[\protect\citeauthoryear{{Scoccimarro}}{{Scoccimarro}}{1998}]{1998MNRAS.299.1097S}
{Scoccimarro} R.,  1998, \mn@doi [\mnras] {10.1046/j.1365-8711.1998.01845.x},
  \href {http://adsabs.harvard.edu/abs/1998MNRAS.299.1097S} {299, 1097}

\bibitem[\protect\citeauthoryear{{Shin}, {Kim}, {Pichon}, {Jeong}  \&
  {Park}}{{Shin} et~al.}{2017}]{2017ApJ...843...73S}
{Shin} J.,  {Kim} J.,  {Pichon} C.,  {Jeong} D.,   {Park} C.,  2017, \mn@doi
  [\apj] {10.3847/1538-4357/aa74b9}, \href
  {http://adsabs.harvard.edu/abs/2017ApJ...843...73S} {843, 73}

\bibitem[\protect\citeauthoryear{{Springel}}{{Springel}}{2005}]{2005MNRAS.364.1105S}
{Springel} V.,  2005, \mn@doi [\mnras] {10.1111/j.1365-2966.2005.09655.x},
  \href {http://adsabs.harvard.edu/abs/2005MNRAS.364.1105S} {364, 1105}

\bibitem[\protect\citeauthoryear{{Springel} et~al.,}{{Springel}
  et~al.}{2005}]{2005Natur.435..629S}
{Springel} V.,  et~al., 2005, \mn@doi [\nat] {10.1038/nature03597}, \href
  {http://adsabs.harvard.edu/abs/2005Natur.435..629S} {435, 629}

\bibitem[\protect\citeauthoryear{{Starobinsky }}{{Starobinsky
  }}{1992}]{1992JETPL..55..489S}
{Starobinsky } A.~A.,  1992, Journal of Experimental and Theoretical Physics
  Letters, \href {http://adsabs.harvard.edu/abs/1992JETPL..55..489S} {55, 489}

\bibitem[\protect\citeauthoryear{{Starobinsky}}{{Starobinsky}}{1980}]{1980PhLB...91...99S}
{Starobinsky} A.~A.,  1980, \mn@doi [Physics Letters B]
  {10.1016/0370-2693(80)90670-X}, \href
  {http://adsabs.harvard.edu/abs/1980PhLB...91...99S} {91, 99}

\bibitem[\protect\citeauthoryear{{Uhlemann} et~al.,}{{Uhlemann}
  et~al.}{2018}]{2018MNRAS.473.5098U}
{Uhlemann} C.,  et~al., 2018, \mn@doi [\mnras] {10.1093/mnras/stx2616}, \href
  {http://adsabs.harvard.edu/abs/2018MNRAS.473.5098U} {473, 5098}

\bibitem[\protect\citeauthoryear{{Zhao}, {Li}  \& {Koyama}}{{Zhao}
  et~al.}{2011}]{2011PhRvD..83d4007Z}
{Zhao} G.-B.,  {Li} B.,   {Koyama} K.,  2011, \mn@doi [\prd]
  {10.1103/PhysRevD.83.044007}, \href
  {http://adsabs.harvard.edu/abs/2011PhRvD..83d4007Z} {83, 044007}

\makeatother
\end{thebibliography}





\bsp	
\label{lastpage}
\end{document}